\documentclass[twocolumn,showpacs,preprintnumbers,amsmath,amssymb]{revtex4}

\usepackage{graphicx}
\usepackage{dcolumn}
\usepackage{bm}

\newcommand{\newc}{\newcommand}
\newc{\lra}{\leftrightarrow}
\newc{\beq}{\begin{equation}}
\newc{\eeq}{\end{equation}}
\newc{\barr}{\begin{eqnarray}}
\newc{\earr}{\end{eqnarray}}
\def\rpm{R_p \hspace{-0.8em}/\;\:}

\begin{document}

\preprint{APS/123-QED}

\title{DEDICATED
SUPERNOVA DETECTION\\
BY A NETWORK OF NEUTRAL CURRENT SPHERICAL TPC'S.}

\author{J.D. Vergados}
\email{vergados@cc.uoi.gr}
\affiliation{RCNP, Osaka University, Ibaraki 567-0047,Japan.}
 \altaffiliation[Physics Department, University of Ioannina, Ioannina, Greece.]\\
\author{Y. Giomataris}
 \email{ioa@hep.saclay.cea.fr}
\affiliation{CEA, Saclay, DAPNIA, Gif-sur-Yvette, Cedex,France}

\date{\today}

\begin{abstract}
Supernova neutrinos can easily be detected by a spherical gaseous TPC detector  measuring very low energy nuclear recoils. The 
expected rates are quite large for a neutron rich target since the neutrino nucleus neutral current interaction yields a  coherent contribution of all neutrons.  As a matter of fact for a 
typical supernova at $10$ kpc, about 1000 events are expected using 
 a spherical detector  of radius 4 m with Xe gas at a pressure of 10 Atm. A world wide network of several such simple,
stable and low cost supernova detectors with a running time of a few centuries is quite feasible.
\end{abstract}

\maketitle

\section{\label{sec:level1} Introduction}

Neutrinos appear to be excellent probes for studying the deep sky. They
travel large distances with the speed of light. 
They can pass through obstacles, without getting distorted on their way and
they are not affected by the presence of magnetic fields. Thus with neutrinos one can see much further than with light With light one cannot observe further than 50 Mpc (1 Mpc=3.3x106 light years). Furthermore the detection of neutrinos 
 reveals information about the source and more specifically about the source interior. Without neutrinos we would probably know nothing about the sun's interior. Thus neutrinos offer a good hope for understanding the early stages of a supernova.
In a typical supernova an energy of about $10^{53}$ ergs is released in the form of neutrinos
\cite{BEACFARVOG},\cite{SUPERNOVA}. These neutrinos
are emitted within an interval of about $10$ s after the explosion and they travel to Earth undistorted, except that,
on their way to Earth, they may oscillate into other flavors. The phenomenon of neutrino oscillations is by
 now established by the observation of
 atmospheric neutrino oscillations \cite{SUPERKAMIOKANDE} interpreted as
 $\nu_{\mu} \rightarrow \nu_{\tau}$ oscillations, as well as
 $\nu_e$ disappearance in solar neutrinos \cite{SOLAROSC}. These
 results have been recently confirmed by the KamLAND experiment \cite{KAMLAND},
 which exhibits evidence for reactor antineutrino disappearance. Thus for traditional detectors
 relying on the charged current
 interactions the precise event rate may depend critically on the specific properties of the neutrinos. The
time integrated spectra in the case of charged current detectors, like the SNO experiment, depend on the neutrino oscillations \cite{TKBT03}.  This, of course, may turn into an advantage for the study of the neutrino properties \cite{BARGER05}.
 An additional problem is the fact that the charged current cross sections depend on the details of the 
 structure of the nuclei involved. 
 
 During the last years various detectors aiming at detecting recoiling nuclei have been developed
in connection with  dark matter 
searches \cite{CDMALL} with thresholds in the $10$ keV region. Recently, however, it has become feasible to detect neutrinos by measuring
 the recoiling nucleus  and employing gaseous detectors with much lower threshold energies. Thus
 one is able to explore the advantages offered by the neutral current interaction, exploring ideas put forward more than a decade  ago \cite{DKLEIN}. This way the deduced neutrino fluxes do not depend on the  neutrino oscillation parameters (e.g. the mixing angles). Even in our case, however, the obtained rates depend on the assumed characteristic temperature for each flavor, see sec. \ref{sec.supernova} . 
Furthermore this interaction, through its vector component, can lead
 to coherence, i.e. an additive contribution of all nucleons in the nucleus. Since the vector  contribution of the
 protons is tiny, the coherence is mainly due to the neutrons of the nucleus.
 
 In this paper we will derive the amplitude for the differential neutrino nucleus coherent cross section. Then
 we will utilize the available information regarding the energy spectrum of supernova neutrinos and estimate
 the expected number of events for all the noble gas targets. We will show that these results can be exploited
by a network of small and relatively cheap spherical TPC detectors placed in various parts of the world 
(for a description of the apparatus see our
earlier work \cite{NOSTOS1}). The operation of such devices as a network will minimize the background 
problems. There is no need to go underground,
but one may have to go sufficiently deep underwater to balance the high pressure of the gas target. Other types
of detectors have also been proposed \cite{MICROPATTERN},\cite{TWOPHASE}.\\
Large gaseous volumes are easily obtained by employing long drift technology (i.e TPC) that can 
provide massive targets by increasing the gas pressure. Combined with an adequate amplifying 
structure and low energy thresholds, a three-dimensional reconstruction of the recoiling particle, electron
 or nucleus, can be obtained. The use of new micropattern detectors and especially the novel Micromegas 
\cite{GIOMA96} provide excellent spatial and time accuracy that is a precious tool for pattern recognition and
 background rejection \cite{CG01},\cite{GORO05}.
The virtue of using such large gaseous volumes and the new high precision microstrip gaseous detectors 
has been recently discussed in a dedicated workshop \cite{WORKSHOP04} and their relevance for low energy neutrino 
physics and dark matter detection has been widely recognized. Such  low-background low-energy threshold 
systems are actually successfully used in the CAST \cite{AALSETH}, the solar axion experiment, 
and are under development 
for several low energy neutrino or dark matter projects \cite{NOSTOS1},\cite{DARKMATTER}.

\section{\label{sec:level4}The NOSTOS detector network}
\begin{figure}[!ht]
\begin{center}
\includegraphics[scale=0.5]{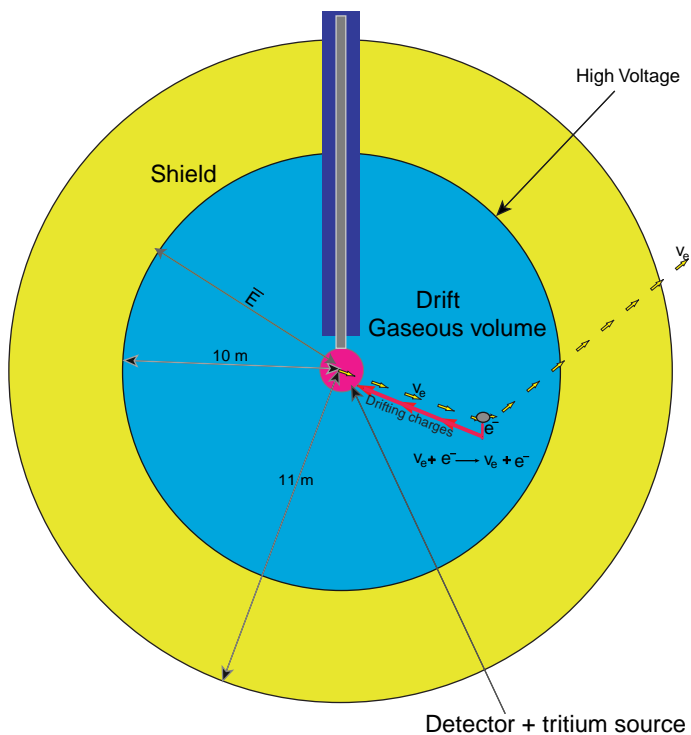}
\caption{A sketch of the NOSTOS detector drawn originally for detecting very low energy electron recoils. The corresponding one for detection of nuclear recoils is analogous.}
 \label{fig:TPC_draw}
 \end{center}
  \end{figure}

\begin{figure}
\begin{center}
\includegraphics[scale=0.4]{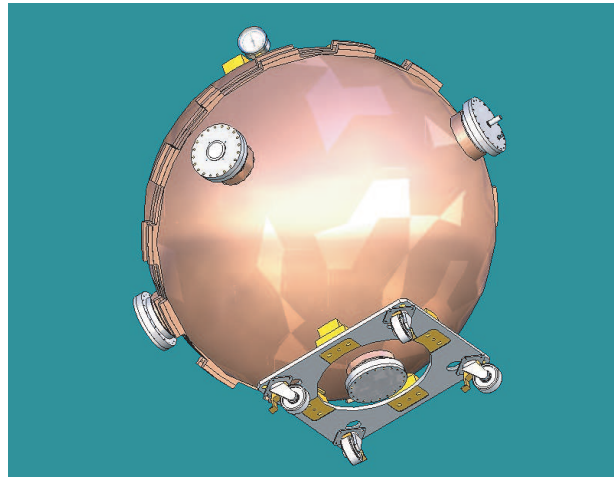}
\caption{First prototype - The SACLAY sphere: $R=1.3~m$, $V=1~m^3$, spherical vessel of Cu $6~mm$ thick, pressure up to $5$ bar (already tested up to $1.5$ bar), vacuum tight, $10^{-6}$ mbar (outgassing $\sim 10^9$ mbar/s)}.
 \label{fig:prototype}
 \end{center}
 \end{figure}

Before we embark on our calculations involving the event rates for supernova neutrino detection of a gaseous 
TPC detector we like to spend a little time in discussing the detector. 
A description of the NOSTOS project and details of the spherical TPC detector are given in \cite{NOSTOS1} and is shown in
\ref{fig:TPC_draw}.
 We have built a spherical prototype 1.3 m in diameter 
which is described in \cite{NOSTOS2}
 (see Fig. \ref{fig:prototype})
. 
The outer vessel is made of pure Cu (6 mm thick)
allowing to sustain pressures up to 5 bar. The inner detector is just a small sphere, 10 mm 
in diameter, made of stainless
steel as a proportional counter located at the center of
curvature of the TPC. We intend to use as amplifying structure
a spherical TPC \cite{GRRC} and  developments are currently under way  
to build a spherical TPC detector using new
technologies. First tests were performed by filling
the volume with argon mixtures and are quite promising. High gains are easily obtained and
 the signal to noise is large enough for sub-keV threshold. The whole system looks stable
 and robust.
The advantages of using the spherical detector concept are the following;
\begin{enumerate}
\item The natural radial focusing of the TPC allows to collect and amplify the deposited
 charges by a simple and robust detector using a single electronic channel to read out
 a large gaseous volume. The small size associated to small detector capacitance permits one
to achieve very low electronic noise. In the present prototype the noise is as low as a 
few hundred electrons and has  easily been obtained; with optimized low noise amplifiers we 
hope to lower it to the level of a few tenths. This is a key point for the obtaining a very
 low energy threshold, i.e. down to 100 eV, by operating the detector at moderate gain of about 100.
 Such low gains are easily obtained at atmospheric pressure and open the way to operate the
 TPC at high pressures. We target to achieve pressures as high as 10 bar for Xenon gas. Even higher pressures
 by a factor 3-6  are aimed at in the case of Argon gas in order to achieve, to first order, the same number
 of events.
\item The radial electric field, inversely proportional to the square of the radius, is a crucial 
point for measuring the depth of the interaction by a simple analysis of the time structure of 
the detector signal. A position resolution of about 10 cm has been already obtained, a fact that is
 of paramount importance for improving the time resolution of the detector and rejecting
 background events by applying fiducial cuts.
\item Building a high pressure metallic sphere, for instance made out of stainless steel or 
copper, seems to assure an excellent quality of the gas mixture and turns out that a single 
gas filling with pure gas is sufficient to maintain the stability of the signal for several 
months. We are pushing the technology to improve the properties of the various elements in
 order to achieve stability over many years.
\item Big high pressure-secure tanks are under development by many international companies
 for hydrogen or oil storage, and therefore the main element of the TPC could be shipped 
at moderate cost.
\end{enumerate}
Our idea is then to  build several such low cost and robust detectors and install them in 
several places over the world.  First estimations show that the required background level
is modest and therefore there is no need for a deep underground laboratory. A mere  100 meter 
water equivalent
coverage seems to be sufficient to reduce the cosmic muon flux at the required level 
(in the case of many such detectors in coincidence, a modest shield is sufficient). 
The maintenance of such system could be easily assured by Universities or even by 
secondary schools. Thanks to the simplicity of the system it could be operated by young 
students with a specific running program and simple maintenance every a few years. 
Notice that such detector scheme, measuring low energy nuclear recoils from neutrino 
nucleus elastic scattering, do not determine the incident neutrino vector and, therefore, 
it is not possible this way to localize the Supernova. A cluster of such detectors in 
coincidence, however, could localize the star by a triangulation technique. \\
 A network of such detectors in coincidence with a sub-keV threshold could also be used o observe
 unexpected low energy events. This low energy range has never been explored using massive
 detectors. A challenge of great importance will be the synchronization of such a detector cluster
with the astronomical $\gamma$-ray burst telescopes to establish whether low energy recoils are
emitted in coincidence with the mysterious $\gamma$ bursts.
\section{\label{sec:level2} Standard and non standard weak interaction}
The standard neutral current  left handed weak interaction can be cast in the form:
\begin{equation}
{\cal L_q}=-\frac{G_F}{\sqrt{2}}
\left[ \bar\nu_\alpha \gamma^\mu (1-\gamma^5)\nu_\alpha \right]
\left[ \bar q\gamma_\mu (g_V(q)-g_A(q)\gamma^5) q \right] 
\label{weak1}
\end{equation}
(diagonal in flavor space) with
\begin{eqnarray}
g_V(u)&=&\frac{1}{2}-\frac{4}{3} \sin^2{\theta_W}~~,~~g_A(u)=\frac{1}{2}~;
\nonumber\\
g_V(d)&=&-\frac{1}{2}+\frac{2 }{3}\sin^2{\theta_W}~~,~~g_A(u)=-\frac{1}{2}
\label{weak2}
\end{eqnarray}
At the nucleon level we get:
\begin{equation}
{\cal L_q}=-\frac{G_F}{\sqrt{2}}
\left[ \bar\nu_\alpha \gamma^\mu (1-\gamma^5)\nu_\alpha \right]
\left[ \bar N\gamma_\mu (g_V(N)-g_A(N)\gamma^5) N \right] 
\label{weak3}
\end{equation}
 with
\begin{eqnarray}
g_V(p)&=&\frac{1}{2}-2 \sin^2{\theta_W}~~,~~g_A(p)=1.27 \frac{1}{2}~;\nonumber\\
g_V(n)&=&-\frac{1}{2}~~,~~g_A(n)=-1.27 \frac{1}{2}
\label{weak4}
\end{eqnarray}
Beyond the standard level one has further interactions which need not be diagonal in flavor space. Thus
\begin{eqnarray}
& &g_V(q)-g_A(q)\gamma^5 \rightarrow 
\nonumber\\
& &\left( g^{SM}_V(q)-g^{SM}_A(q)\gamma^5 \right)
\delta_{\alpha \beta}+\left( \lambda^{qL}\delta_{\alpha \beta}+\epsilon^{qL}_{\alpha \beta} \right)(1-\gamma^5)
\nonumber\\
& &\left[ \bar\nu_\alpha \gamma^\mu (1-\gamma^5)\nu_\alpha \right]\rightarrow
\left[ \bar\nu_\alpha \gamma^\mu (1-\gamma^5)\nu_\beta \right]
\label{weak5}
\end{eqnarray}
Furthermore at the nucleon level
\begin{eqnarray}
g_V(N)-g_A(N)\gamma^5 \rightarrow& & 
\left( g^{SM}_V(N)-g^{SM}_A(N)\gamma^5 \right) \delta_{\alpha \beta}
\nonumber\\
& & + \left( \lambda^{NL}\delta_{\alpha \beta}+\epsilon^{NL}_{\alpha \beta} \right)(1-1.27 \gamma^5)
\nonumber\\
\left[ \bar\nu_\alpha \gamma^\mu (1-\gamma^5)\nu_\alpha \right]\rightarrow & &
\left[ \bar\nu_\alpha \gamma^\mu (1-\gamma^5)\nu_\beta \right]
\label{weak6}
\end{eqnarray}
with
\begin{eqnarray}
\lambda^{pL}&=&2 \lambda^{uL}+\lambda^{dL}~,~\lambda^{nL}=\lambda^{dL}+ 2\lambda^{dL}~,
\nonumber\\
& &\epsilon^{pL}_{\alpha \beta}=2\epsilon^{uL}_{\alpha \beta}+\epsilon^{dL}_{\alpha \beta}~,
~\epsilon^{nL}_{\alpha \beta}=\epsilon^{uL}_{\alpha \beta}+ 2\epsilon^{dL}_{\alpha \beta}
\label{weak7}
\end{eqnarray}
In the above expressions $\lambda^{qL}$ can arise, e.g.,  from radiative corrections, see e.g. PDG \cite{PDG} and $\epsilon^{qL}_{\alpha \beta}$ from R-parity violating interactions in supersymmetric models \cite{HIRSCH00}-\cite{HFVK00}.\\
Indeed since R-parity conservation has no robust
theoretical motivation one may accept an extended framework of
the MSSM with R-parity non-conservation  MSSM.
In this case the superpotential
$W$ acquires additional R-paity violating terms:
\begin{eqnarray}
  W_{\rpm} & = & \lambda_{ijk}L_{i}L_{j}E^c_{k}
  + \lambda^{\prime}_{ijk}L_{i}Q_{j}D^c_{k}\nonumber\\ 
 &+& \lambda^{\prime \prime}_{ijk} U^c_{i}D^c_{j}D^c_{k}+ \mu_j L_{j}H_u
    \label{rpsuperpotential}
\end{eqnarray}
Of interest to us here is the $\lambda^{\prime}_{ijk}L_{i}Q_{j}D^c_{k}$ involving first generation quarks and s-quarks, 
i.e the term $\lambda^{\prime}_{\alpha 1 1} L_{\alpha}Q_{1}D^c_{1}$. From this term in four component notation we get the contribution
$$\lambda^{\prime}_{\alpha 1 1} \left (\bar{d^c_R} \nu_{\alpha L}- \bar{u^c_R} \alpha_L \right )\tilde{d}^c,~\alpha =e,\mu,~\tau$$
where $\nu_{\alpha L}=\frac{1}{2}(1-\gamma_5)\nu_{\alpha}$ etc. Thus 
\begin{itemize}
\item The first term at tree level yields the interaction
\beq
 -\frac{\lambda^{\prime}_{\alpha 1 1} \lambda^{\prime}_{\beta 1 1}}{m^2_{\tilde{d}_L}} 
 \bar{\nu}_{\alpha L} d^c_R  \bar{d^c_R} \nu_{\beta L}
\label{rparity1}
\eeq
By performing a Fierz transformation we can rewrite it in the form:
\beq
 \frac{1}{2}\frac{\lambda^{\prime}_{\alpha 1 1} \lambda^{\prime}_{\beta 1 1}}{m^2_{\tilde{d}_L}} 
 \bar{\nu}_{\alpha L} \gamma^{\mu} \nu_{\beta L} \bar{d^c_R} \gamma_{\mu} d^c_R
\label{rparity2}
\eeq
The previous equation can be cast in the form:
\begin{eqnarray}
{\cal L}_d&=&-\frac{G_F}{\sqrt{2}} \epsilon ^d_{\alpha \beta}
\left[ \bar{\nu_{\alpha}} \gamma^\mu ( 1-\gamma^5 )\nu_\beta \right]\left[ \bar{d}\gamma_\mu ( 1-\gamma^5 )d \right ];
\nonumber\\
 \epsilon ^d_{\alpha \beta}&=& \lambda^{\prime}_{\alpha 1 1} \lambda^{\prime}_{\beta 1 1} \frac{m_W^2}{m^2_{\tilde{d}_L}} 
\label{rparity3}
\end{eqnarray}
There is no such term associated with the $u$ quark, $\epsilon ^u_{\alpha \beta}=0$.
\item Proceeding in an analogous fashion the collaborative effect  of the first and second term, for $\alpha,\beta=e,\mu,\tau$, yields
the charged current contribution:
\begin{eqnarray}
{\cal L}_{du}&=&\frac{G_F}{\sqrt{2}} \epsilon ^d_{\alpha \beta}
\left[ \bar{\alpha} \gamma^\mu (1-\gamma^5)\nu_\beta \right]\left[ \bar{u}\gamma_\mu (1-\gamma^5)d \right ]
\label{rparity4}
\end{eqnarray}
\item Finally the second term, for $\alpha,\beta=e,\mu,\tau$, leads to a neutral current contribution of the charged leptons:
\begin{eqnarray}
{\cal L}_u &=&\frac{G_F}{\sqrt{2}} \epsilon ^d_{\alpha \beta}
\left[ \bar{\alpha} \gamma^\mu ( 1-\gamma^5 ) \beta \right]\left[ \bar{u}\gamma_\mu ( 1-\gamma^5 )u \right ]
\label{rparity5}
\end{eqnarray}
\end{itemize}
The above non standard flavor changing neutral current interaction have been found to play an important role in the in the infall stage
of a stellar collapse \cite{AFG05}. Furthermore precise measurements involving the neutral current neutrino-nucleus interactions may yield valuable information about the non standard interactions \cite{BMR05}. They are not, however, going to be further considered in this work.
 
\section{\label{sec:level3} Elastic Neutrino nucleon Scattering}
\begin{figure}[!ht]
\begin{center}
 \rotatebox{90}{\hspace{0.0cm} {$\rightarrow \frac{T^{max}_{recoil}}{m_{recoil}}$}}
\includegraphics[scale=0.7]{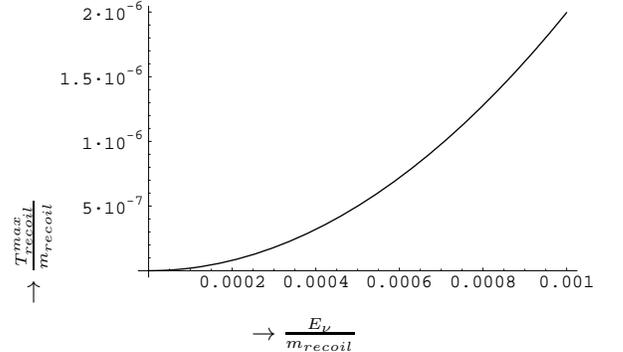}
\hspace{6.0cm}$\rightarrow \frac{E_{\nu}}{m_{recoil}}$
\caption{The maximum recoil energy  as a function of the neutrino energy (both in units of the
recoiling mass). The scale is realistic for nuclear recoils of the type of experiments considered here.}
 \label{fig:yofx}
 \end{center}
  \end{figure}

\begin{figure}[!ht]
 \begin{center}
 \rotatebox{90}{\hspace{1.0cm} {$\rightarrow \frac{E^{min}_{\nu}}{m_{recoil}}$}}
\includegraphics[scale=0.7]{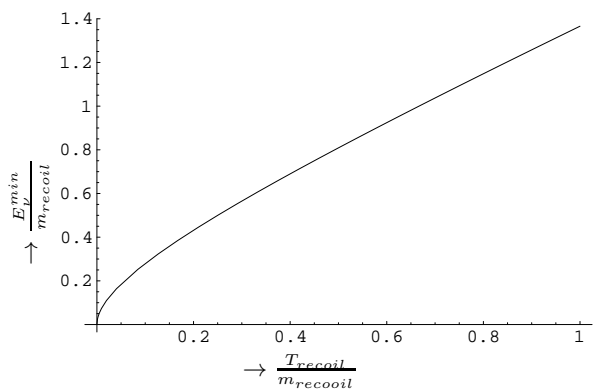}
\hspace{6.0cm}$\rightarrow \frac{T_{recoil}}{m_{recooil}}$
 \caption{The minimum neutrino energy as a function of the recoil energy (both in units of the
recoiling mass). The scale is not realistic for nuclear recoils considered here, but relevant for other
experiments. The realistic scale for experiments considered here can be deduced from Fig.
\ref{fig:yofx} by exchanging the coordinate axes.}
 \label{fig:xofy}
 \end{center}
  \end{figure}

The cross section for elastic neutrino nucleon scattering has extensively been studied.
It has been shown that at low energies the weak differential cross section can be simplified and  be cast in the form:
\cite{BEACFARVOG},\cite{VogEng}:
 \begin{eqnarray}
 \left(\frac{d\sigma}{dT_N}\right)_{w}&=&\frac{G^2_F m_N}{2 \pi}
 [(g_V+g_A)^2\\
\nonumber
&+& (g_V-g_A)^2 [1-\frac{T_N}{E_{\nu}}]^2
+ (g_A^2-g_V^2)\frac{m_NT_N}{E^2_{\nu}}]
 \label{elasw}
  \end{eqnarray}
  where $m_N$ is the nucleon mass and $g_V$, $g_A$ are the weak coupling constants. Neglecting their
  dependence on the momentum transfer to the nucleon they take the form (see previous section):
  \beq
 g_V=-2\sin^2\theta_W+1/2\approx 0.04~,~g_A=\frac{1.27}{2} ~~,~~(\nu,p)
\label{gcoup1}
\eeq
\beq
g_V=-1/2~,~g_A=-\frac{1.27}{2}~~,~~(\nu,n)
 \label{gcoup2}
 \eeq
 In the above expressions for the axial current the renormalization
in going from the quark to the nucleon level was taken into account. For antineutrinos $g_A\rightarrow-g_A$.
To set the scale we write:
\beq \frac{G^2_F m_N}{2 \pi}=5.14\times 10^{-41}~\frac{cm^2}{MeV}
\label{weekval} 
\eeq
The nucleon energy depends on the  neutrino energy and the
scattering angle, the angle between the direction of the recoiling particle and that of the incident neutrino. In the laboratory frame  it is given by:
\beq
T_N= \frac{2~m_N (E_{\nu}\cos{\theta})^2}{(m_N+E_\nu)^2-(E_{\nu}
\cos{\theta})^2}~~,~~0\leq \theta\leq \pi/2
\label{eq:TN}
\eeq
(forward scattering). For sufficiently small neutrino energies, the last equation can be simplified as follows:
$$ T_N \approx \frac{ 2(E_\nu \cos{\theta})^2}{m_N}$$
The above formula can be generalized to any target and can be written in dimensionless form
as follows:
\beq
y=\frac{2\cos^2{\theta}}{(1+1/x_{\nu})^2-\cos^2{\theta}}~~,~~
y=\frac{T_{recoil}}{m_{recoil}},x_{\nu}=\frac{E_{\nu}}{m_{recoil}}
\label{recoilen}
\eeq 
 The maximum  energy occurs when $\theta=0$, $y_{max}=\frac{2}{(1+1/x_{\nu})^2-1}$,
in agreement with Eq. (2.5) of ref. \cite{BEACFARVOG}. This relationship is
plotted in  Fig. \ref{fig:yofx}.
  One can invert Eq. \ref{recoilen} and get the  neutrino energy associated with a given recoil energy and
scattering angle. One finds
\beq
  x_{\nu}=\left[-1+\cos{\theta} \sqrt{1+\frac{2}{y}} \right]^{-1}~~,~~0\leq \theta\leq \pi/2
  \label{xofyxi}
  \eeq
  The minimum neutrino energy for a give recoil energy is given by:
    \beq
  x^{min}_{\nu}=\left[-1+\sqrt{1+\frac{2}{y}} \right]^{-1}=\frac{y}{2}(1+\sqrt{1+\frac{2}{y}})
  \label{xofy}
  \eeq
 in agreement with Eq. (4.2) of ref. \cite{BEACFARVOG}. The last equation is useful in obtaining the differential cross section (with respect to the recoil energy) after folding with the neutrino spectrum
  and it is shown in Fig. \ref{fig:xofy}
\section{\label{sec:level5}Coherent neutrino nucleus scattering}
\begin{figure}[!ht]
 \begin{center}
 \rotatebox{90}{\hspace{1.0cm} {\tiny {$F_{coh}$}}}
\includegraphics[scale=0.6]{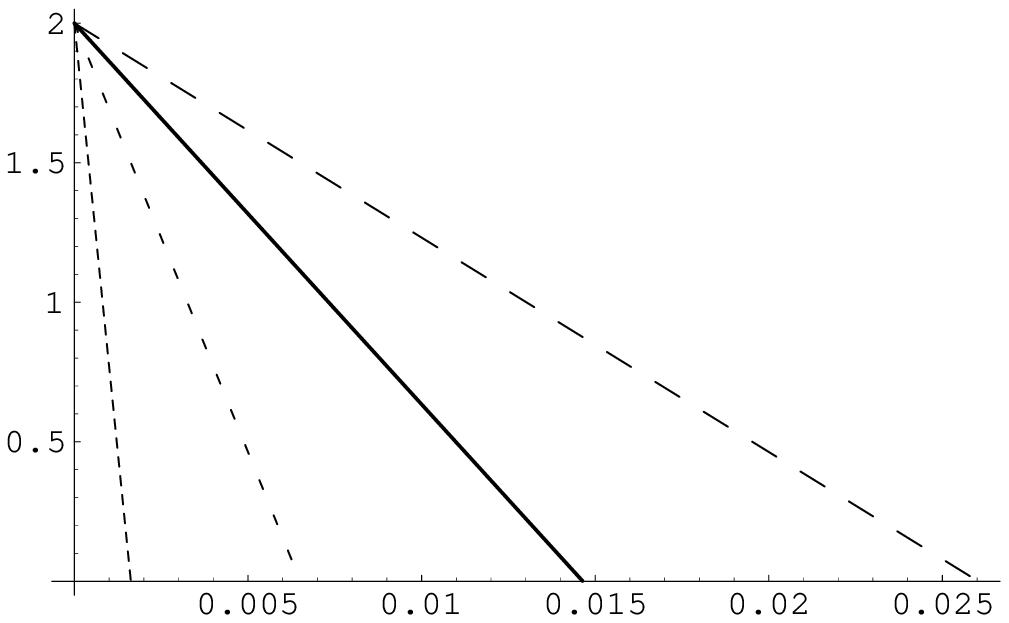}
\hspace*{0.0cm}\tiny{$T_A \rightarrow$ MeV}\\
 \rotatebox{90}{\hspace{1.0cm} {\tiny {$F_{coh}$}}}
\includegraphics[scale=0.6]{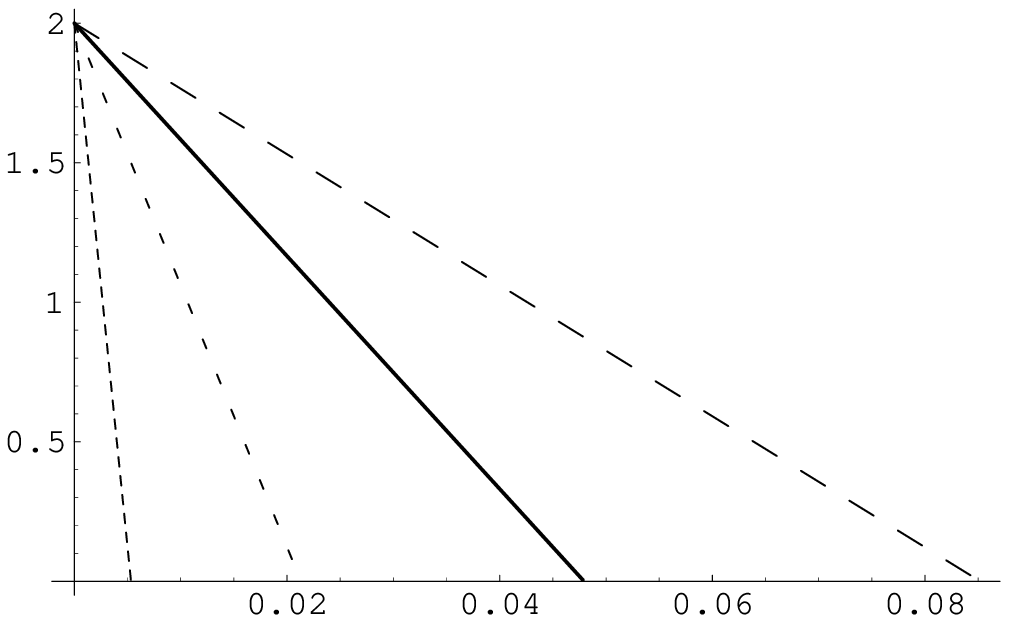}
\hspace*{0.0cm}\tiny{$T_A \rightarrow$ MeV}
 \caption{The function $F_{coh}(A,T_A,E_{\nu})$ as a function of the recoil energy $T_A$ for,
from left to right, $E_{\nu}=10,~20,~30,~40$
MeV. The results shown are for Xe on the top and Ar at the bottom}
 \label{fig:fcoh}
 \end{center}
  \end{figure}

From the above expressions we see that the vector current contribution, which may lead to coherence, is negligible
in the case of the protons. Thus the coherent contribution \cite{PASCHOS} may come from the neutrons and is expected to be
proportional to the square of the neutron number.
The neutrino-nucleus scattering can be obtained from the amplitude of the neutrino nucleon scattering under
the following assumptions:
\begin{itemize}
\item Employ the appropriate kinematics, i.e. those involving the elastically scattered nucleus.
\item Ignore  effects of the nuclear form factor. Such effects, which are not expected to be very large,
 are currently under study and they will
 appear elsewhere.
\item The effective neutrino-nucleon amplitude is obtained as above with the substitution 
$${\bf q}\Rightarrow \frac{{\bf p}}{A}~~,~~E_N \Rightarrow \sqrt{m_N^2+\frac{{\bf p}^2}{A^2}}=\frac{E_A}{A}$$
with ${\bf q}$ the nucleon momentum and ${\bf p}$ the nuclear momentum.  
\end{itemize}
Under the above assumptions the neutrino-nucleus cross section takes the form:
 \begin{eqnarray}
 \left(\frac{d\sigma}{dT_A}\right)_{w}&=&\frac{G^2_F Am_N}{2 \pi} \times 
 \nonumber\\
&& [(M_V+M_A)^2 \left (1+\frac{A-1}{A}\frac{T_A}{E_{\nu}} \right )
 \nonumber\\
+ (M_V-M_A)^2 
(1&-&\frac{T_A}{E_{\nu}})^2
\left (1-\frac{A-1}{A}\frac{T_A}{m_N}\frac{1}{E_{\nu}/T_A-1} \right )
\nonumber\\
&+& (M_A^2-M_V^2)\frac{Am_NT_A}{E^2_{\nu}} ]
 \label{elaswA}
  \end{eqnarray}
  Where $M_V$ and $M_A$ are the nuclear matrix elements associated with the vector and the axial current
  respectively and $T_A$ is the energy of the recoiling nucleus.
 The axial current contribution vanishes for $0^+ \Rightarrow 0^+$ transitions. Anyway it is negligible
  in front of the coherent scattering due to neutrons. Thus the previous formula is reduced to:
   \begin{eqnarray}
 & &\left(\frac{d\sigma}{dT_A}\right)_{w}=\frac{G^2_F Am_N}{2 \pi}~(N^2/4)F_{coh}(A,T_A,E_{\nu}),
\nonumber\\
& &F_{coh}(A,T_A,E_{\nu})=
  \left (1+\frac{A-1}{A}\frac{T_A}{E_{\nu}} \right )
+(1-\frac{T_A}{E_{\nu}})^2
\nonumber\\
& &\left (1-\frac{A-1}{A}\frac{T_A}{m_N}\frac{1}{E_{\nu}/T_A-1} \right )
-\frac{Am_NT_A}{E^2_{\nu}} 
 \label{elaswAV}
  \end{eqnarray}
  The function $F_{coh}(A,T_A,E_{\nu})$ is shown in Fig \ref{fig:fcoh} as a function of the recoil
  energy in the case of Ar (N=22) and Xe (N=77) for $10,20,30$ and $40$ MeV respectively.  
  We see two reasons for enhancement of the cross section: 
\begin{itemize}
\item The overall A factor due to the kinematics, which is  counteracted by the smaller
nuclear recoil energy when compared to the nucleon recoil energy for the same neutrino energy. 
This factor will be absorbed into the energy integrals, see the function $F_{fold}(A,T,(T_A)_{th})$ below.
\item The $N^2$ enhancement due to coherence.
\end{itemize}
\section{\label{sec.supernova} Supernova neutrinos}
\begin{figure}[!ht]
 \begin{center}
 \rotatebox{90}{\hspace{1.0cm} {$\frac{dN}{d E_{\nu}}\rightarrow \frac{10^{58}}{MeV}$}}
\includegraphics[scale=0.7]{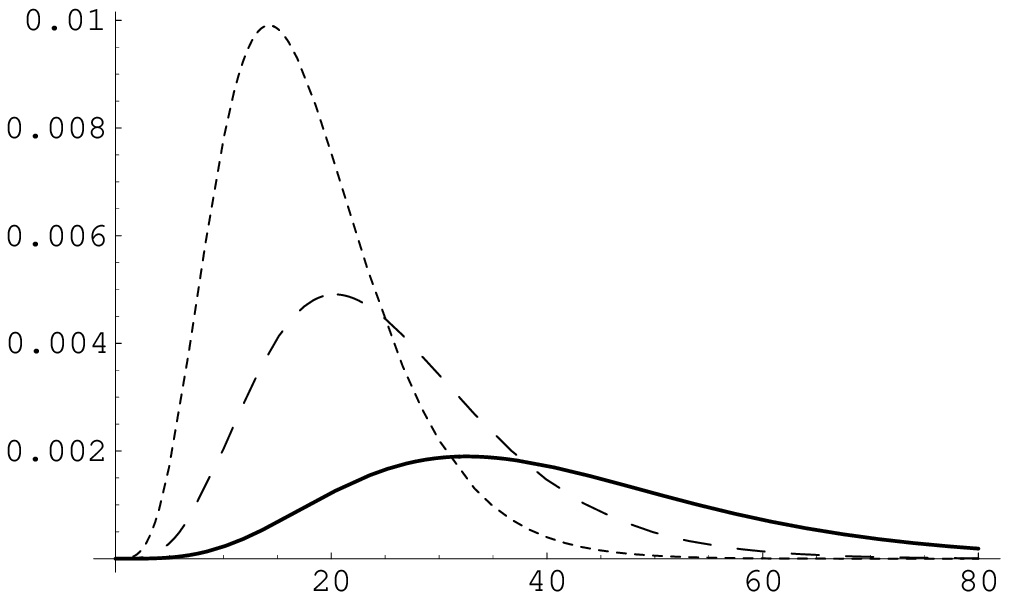}
\hspace{8.0cm}$E_{\nu} \rightarrow$ MeV
 \caption{The supernova neutrino spectrum. The short dash, long dash and continuous curve correspond
to $\nu_e,\tilde{\nu}_e$ and all other flavors respectively}
 \label{supernovasp}
 \end{center}
  \end{figure}

\begin{figure}[!ht]
 \begin{center}
\includegraphics[scale=0.7]{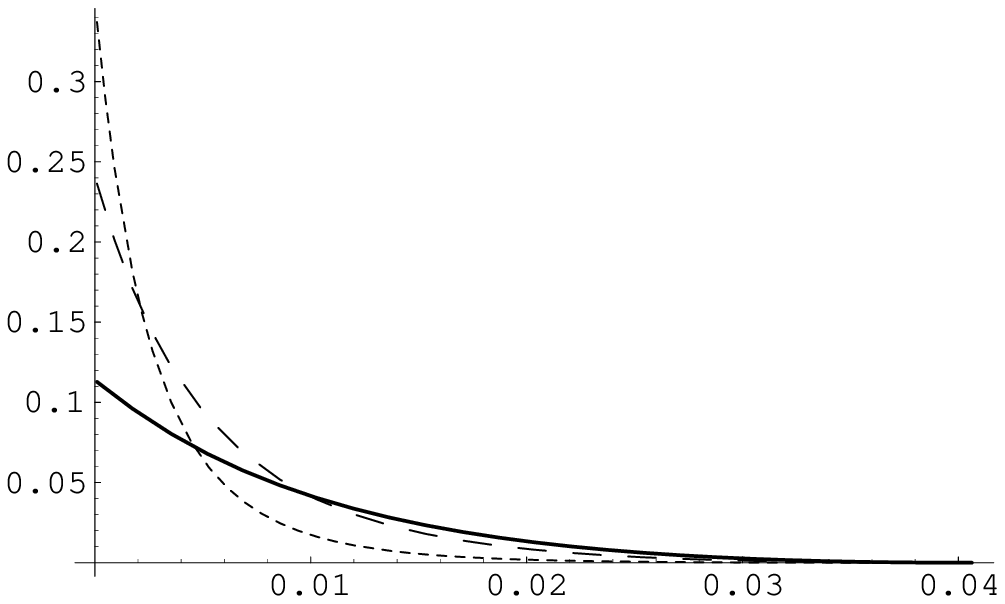}
\hspace*{0.0cm}\tiny{$T_A \rightarrow$ MeV}\\
\includegraphics[scale=0.7]{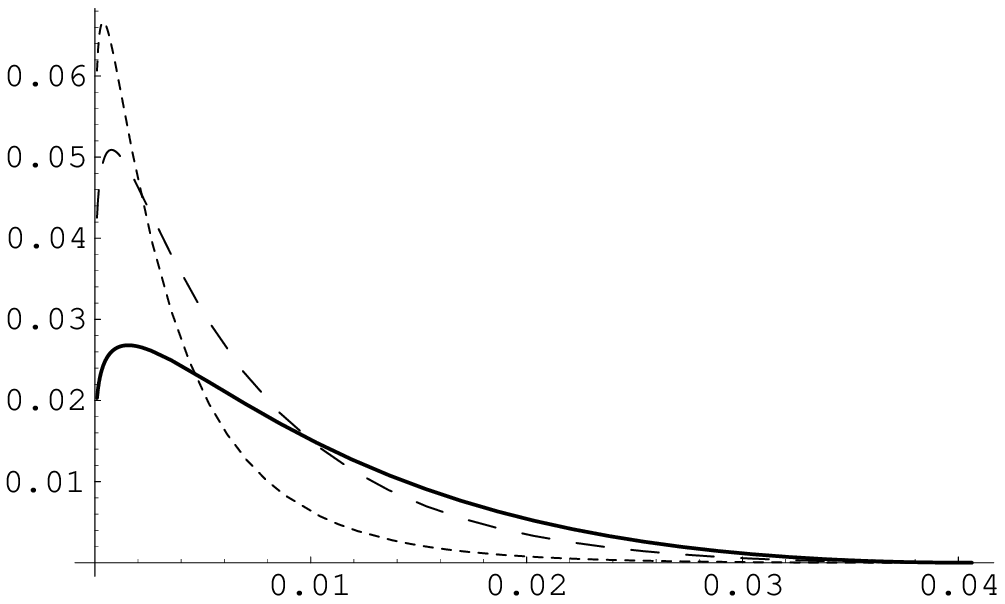}
\hspace*{0.0cm}\tiny{$T_A \rightarrow$ MeV}
 \caption{The differential event rate as a function of the recoil energy $T_A$, in arbitrary units, for
Xe. On  the top we show the results without quenching, while at the bottom the quenching factor is
included. We notice that the effect of quenching is more prevalent at low energies. The notation 
for each neutrino species is 
the same as in Fig. \ref{supernovasp}}
 \label{fig:difr131}
 \end{center}
  \end{figure}

The number of neutrino events for a given detector depends on the neutrino spectrum and the distance of the
source. We will consider a typical case of a source which is about $10$ kpc, l.e. $D=3.1 \times 10^{22}$ cm ( of the order of the radius of the galaxy) with 
an energy output of $3 \times 10^{53}$ ergs with a duration of about $10$ s.\\
The neutrino spectra are parametrized as follows:
\beq
\frac{dN}{dE_{\nu _i}}=\frac{\Phi_i}{\prec E_i\succ }\frac{\beta_i^{\beta_i}}{\Gamma(\beta_i}
 \left(\frac{E_i}{\prec E_i \succ}  \right)^{\beta_i-1} Exp \left(-\beta_i \frac{E_i}{\prec E_i \succ} \right)
 \label{nudistr1}
 \eeq
 where $\prec E_i \succ $ is the average energy of neutrino flavor i with $E_{\nu_e}<E_{\bar{\nu}_e}<E_{\nu_x}(E_{\bar{\nu}_x})$ for $x=\mu,\tau$. The parameters $\Phi_i,~\prec E_i \succ $ and $\beta_i$ for each flavor, $\nu_e,\bar{\nu}_e,\nu_x(\bar{\nu}_x)$ with $x=\mu,\tau$ are determined phenomenologically \cite{NUSPECTRA}.   In the present paper, in order to minimize the number of parameters, we will assume for simplicity that each neutrino flavor is characterized by  a 
 Fermi-Dirac like distribution times its characteristic cross section, which is adequate for our purposes. Thus: 
\beq
\frac{dN}{dE_{\nu}}=\sigma(E_{\nu})\frac{E^2_{\nu}}{1+exp(E_{\nu}/T)}=\frac{\Lambda}{JT}\frac{x^4}{1+e^x}~~,
~~x=\frac{E_{\nu}}{T}
\label{nudistr}
\eeq
with
$J=\frac{31\pi^6}{252}$, $\Lambda$  a constant and 
$T$ the temperature of the emitted neutrino flavor. 
Each flavor is characterized by its
own temperature as follows:
$$T=8 \mbox { MeV for } \nu_{\mu},\nu_{\tau},\tilde{\nu}_{\mu}, \tilde{\nu}_{\tau}
\mbox{,} T=5 ~(3.5)\mbox{ MeV for } \tilde{\nu}_e ~(\nu_e)$$
The constant $\Lambda$ is determined by the requirement that the distribution yields the total energy of each
neutrino species.
$$U_{\nu}=\frac{\Lambda T}{J}\int_0^{\infty } dx \frac{x^5}{1+e^x}\Rightarrow \Lambda=\frac{U_{\nu}}{T}$$
We will further assume that  $U_{\nu}=0.5 \times 10^{53}$ ergs
per neutrino flavor. Thus one finds:
$$\Lambda=0.89\times 10^{58}~(\nu_e),~~0.63\times 10^{58}~(\tilde{\nu}_e)~,0.39\times 10^{58}
\mbox{ ( other)}$$
The emitted neutrino spectrum is shown in Fig. \ref{supernovasp}.
  The differential event rate (with respect to the recoil energy) is proportional to the quantity:
\beq
\frac{dR}{dT_A}=\frac{\lambda (T)}{J}\int_0^{\infty } dx
F_{coh}(A,T_A,xT) \frac{x^4}{1+e^x}
\label{dRdT}
\eeq
with $\lambda(T)=(0.89,0.63,0.39)$  for $\nu_e,\tilde{\nu}_e$ and all other flavors respectively. 
This is shown  in Figs. \ref{fig:difr131} and \ref{fig:difr131}.
   \begin{figure}[!ht]
 \begin{center}
\includegraphics[scale=0.7]{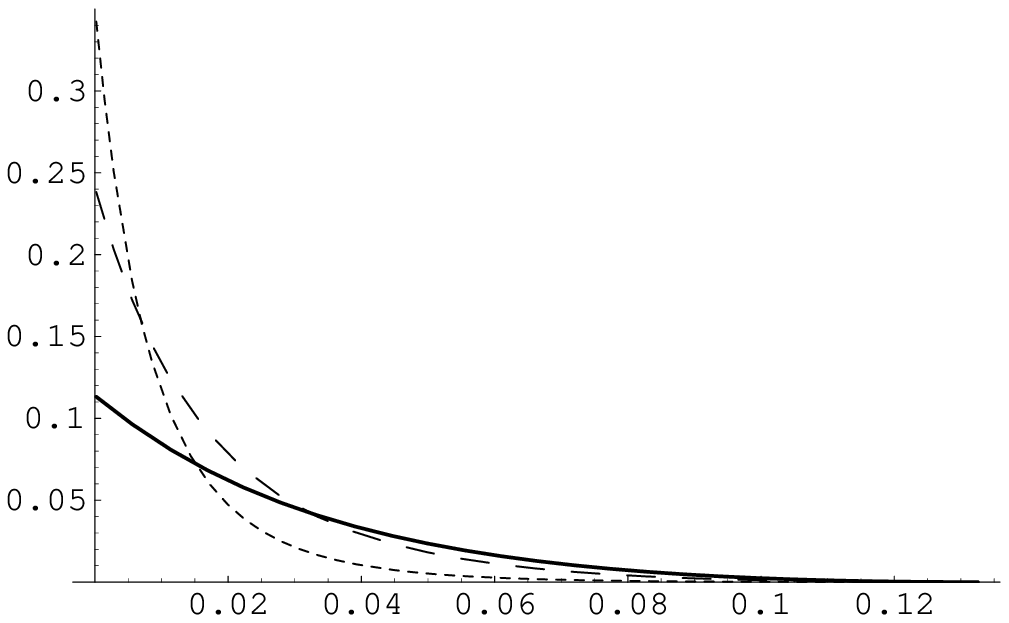}
\hspace*{0.0cm}\tiny{$T_A \rightarrow$ MeV}\\
\includegraphics[scale=0.7]{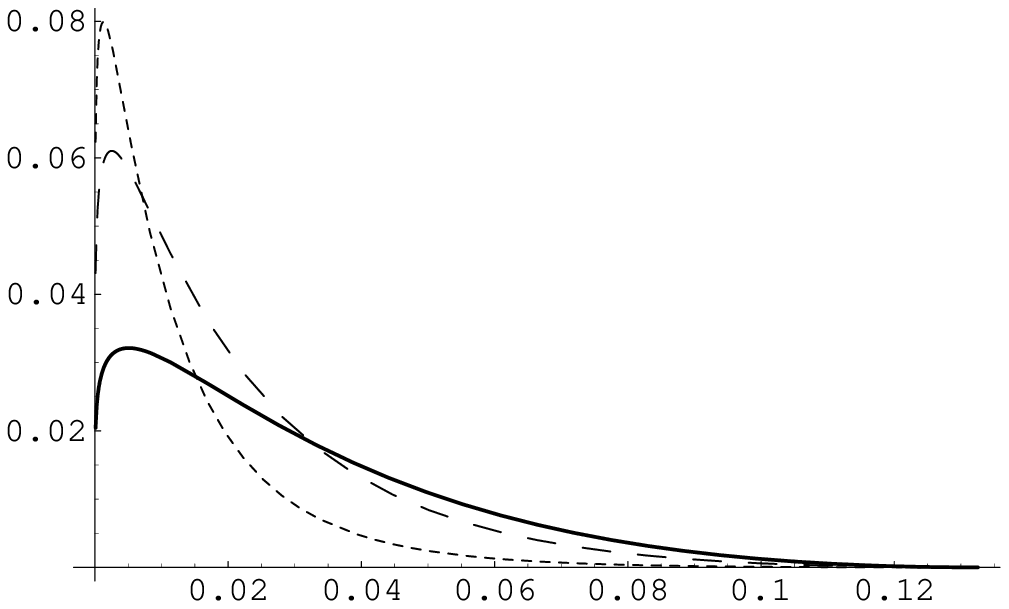}
\hspace*{0.0cm}\tiny{$T_A \rightarrow$ MeV}
 \caption{The same as in Fig. \ref{fig:difr131} for the Ar target.}
 \label{fig:difr40}
 \end{center}
  \end{figure}
The total number of expected events for each neutrino species can be cast in the form:
\begin{eqnarray}
\mbox{No of events}&=&\tilde{C}_{\nu} (T) h(A,T,(T_A)_{th}),
\nonumber\\
\\h(A,T,(T_A)_{th})
&=&\frac{F_{fold}(A,T,(T_A)_{th})}{F_{fold}(40,T,(T_A)_{th})}
\label{events}
\end{eqnarray}
with
\begin{eqnarray}
F_{fold}(A,T,(T_A)_{th})&=&\frac{A}{J}\int_{(T_A)_{th}}^{(T_A)_{max}}\frac{dT_A}{1MeV}\times 
\nonumber\\ 
\int_0^{\infty }&& dx F_{coh}(A,T_A,xT) \frac{x^4}{1+e^x}
\label{events1}
\end{eqnarray}
and
\beq
\tilde{C}_{\nu}(T)=\frac{G^2_F m_N1MeV}{2 \pi}
 \frac{N^2}{4}\Lambda (T)\frac{1}{4 \pi D^2}\frac{PV}{kT_0}
\label{C1}
\eeq
Where $k$ is Boltzmann's constant, $P$ the pressure, $V$ the volume, and $T_0$ the temperature of the gas.

Summing over all the neutrino species we can write:
\beq
\mbox{No of events}=C_{\nu} r(A)\frac{K(A,(T_A)_{th})}{K(40,(T_A)_{th})}Qu(A)
\label{sumevents}
\eeq
with
\begin{eqnarray}
&&C_{\nu}=153  \left ( \frac{N}{22} \right )^2 \times 
\nonumber\\
&&\frac{U_{\nu}}{0.5\times 10^{53}ergs}
\left ( \frac{10kpc}{D}\right )^2 \frac{P}{10Atm}
\left[ \frac{R}{4m}\right]^3 \frac{300}{T_0}
\label{C2}
\end{eqnarray}
In the above expression $r(A)$ is a kinematical parameter depending on the nuclear mass number,
which is essentially unity.

\begin{figure}[!ht]
 \begin{center}
 \rotatebox{90}{\hspace{-2.0cm} {$r(A)\rightarrow $}}
 \hspace{8.0cm}$A \rightarrow$ 
\includegraphics[scale=0.7]{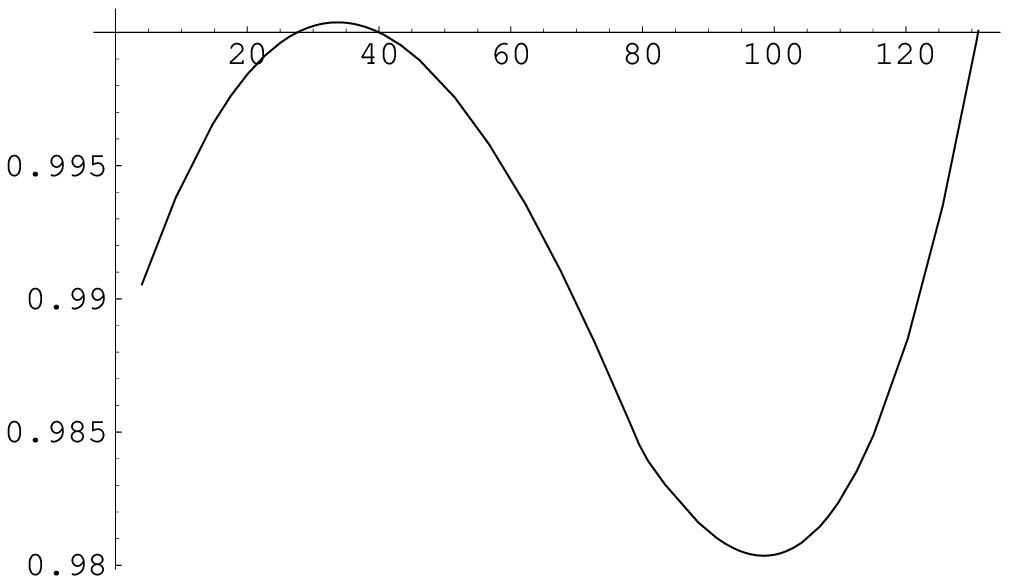}
 \caption{The function $r(A)$ versus the nuclear mass number. To a good approximation $r(A)\simeq 1.0$
(for definitions see text)}
 \label{fig:ratio}
 \end{center}
  \end{figure}

\begin{figure}[!ht]
 \begin{center}
 \rotatebox{90}{\hspace{-0.0cm} {${\tiny K(A,(T_A)_{th})}\rightarrow $}}
\includegraphics[scale=0.7]{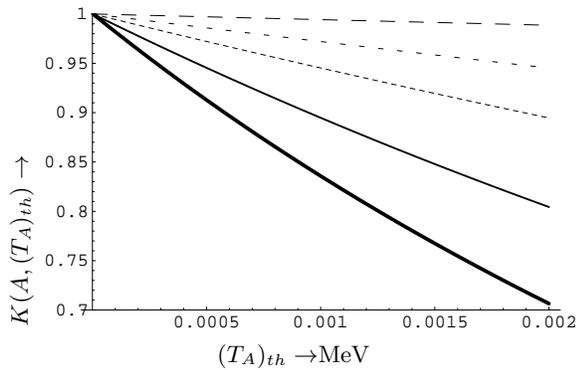}
 \hspace{1.0cm}${\tiny (T_A)_{th}} \rightarrow$MeV 
 \caption{The function $K(A,(T_A)_{th})$ versus $(T_A)_{th}$ for various  nuclear mass numbers
without the quenching factor.
From  top to bottom He, Ne, Ar, Kr and Xe.
(for definitions see text)}
 \label{fig:K}
 \end{center}
  \end{figure}

\begin{figure}[!ht]
 \begin{center}
 \rotatebox{90}{\hspace{-0.0cm} {\tiny $K(A,(T_A)_{th})\rightarrow $}}
\includegraphics[scale=0.75]{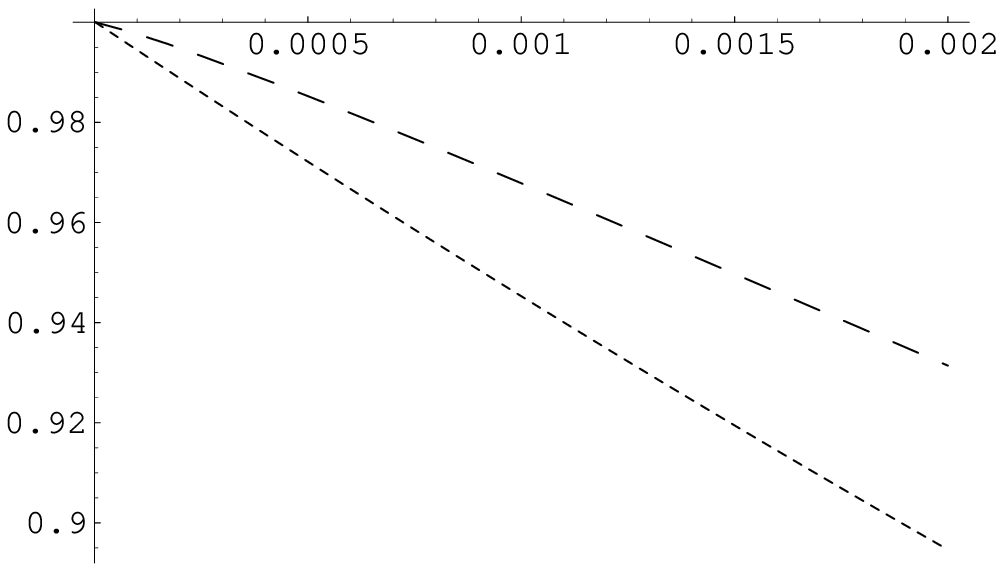}
 \hspace*{0.0cm}\tiny{$(T_A)_{th} \rightarrow$MeV }\\
  \rotatebox{90}{\hspace{-0.0cm} {\tiny $K(A,(T_A)_{th})\rightarrow $}}
\includegraphics[scale=0.75]{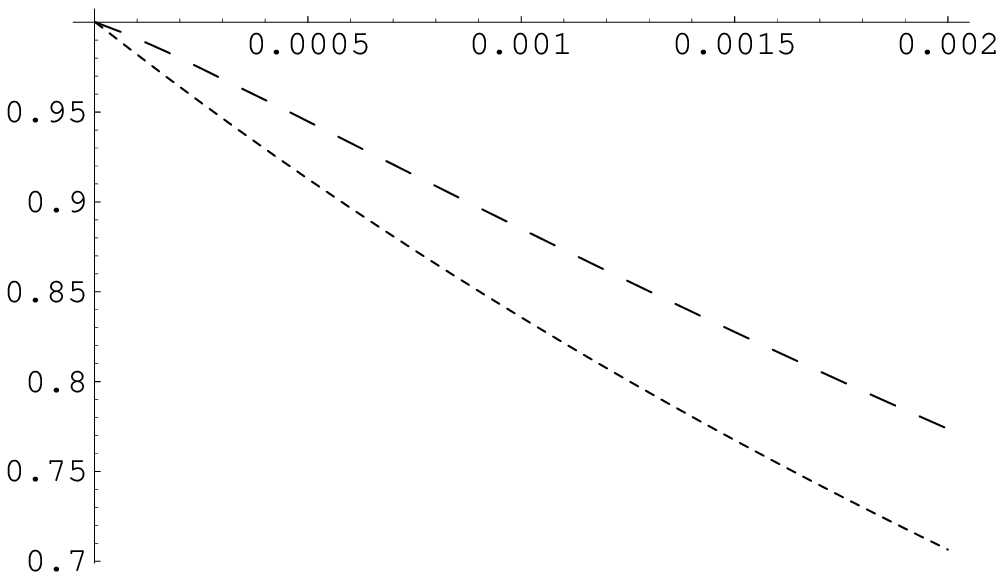}
\hspace*{0.0cm}\tiny{$(T_A)_{th} \rightarrow$MeV }
 \caption{The function $K(A,(T_A)_{th})$ versus $(T_A)_{th}$ for the target Ar on the top and
Xe at the bottom. The short and long dash correspond to no quenching and quenching factor respectively.
One sees that the effect of quenching is less pronounced at higher thresholds.
The differences appear small, since we present here only  the ratio of the rates to that at zero
threshold. The effect of quenching at some specific threshold energy is not shown here.
For a threshold energy of $100$ eV the rates are quenched by factors of $3$ and $3.5$ for Ar and Xe
 respectively (see Eq. (\ref{quench2}).}
 \label{fig:Kqu}
 \end{center}
  \end{figure}

 $K(A,(T_A)_{th})$
 is  the rate at a given threshold energy divided by that at zero threshold. It depends
 on the threshold
energy, the assumed quenching factor and the nuclear mass number. It is unity at $(T_A)_{th})=0$.
The function $r(A)$ is plotted in \ref{fig:ratio}. It is seen that  it can be well approximated by unity.\\
 From the above equation we find
that, ignoring quenching, the following expected number of events:
\beq
1.25,~31.6,~153,~614,~1880\mbox{ for He, Ne, Ar, Kr and Xe}
\label{allrates}
\eeq
respectively. For other possible targets the rates can be found by the above formulas or interpolation.\\
  The function $K(A,(T_A)_{th})$ is plotted in Fig.
\ref{fig:K} for threshold energies up to $2$keV. 
  We see that the threshold effects are stronger in heavier systems since, on the average, the transfered
  energy is smaller. Thus for a threshold energy of $2$ keV in the case of Xe the number of events is reduced
  by $30\%$ compared to those at zero threshold.\\
 The quantity $Qu(A)$ is a factor less than one multiplying the total rate, assuming a  threshold energy
  $(T_A)_{th}=100$eV, due to the quenching. The idea of quenching is introduced, since, for low emery recoils,
 only a fraction of the total deposited energy goes into
 ionization. The ratio of the amount of ionization induced in the gas due to nuclear recoil to the amount of ionization
 induced 
by an electron of the same kinetic energy is referred to as a quenching factor $Q_{fac}$. This factor depends mainly on the 
detector material, the recoiling energy as well as the process considered \cite{SIMON03}.
 In our estimate of $Qu(T_A)$ we assumed a quenching factor of the following empirical form motivated by the Lidhard
theory \cite{SIMON03}-\cite{LIDHART}:
\beq
Q_{fac}(T_A)=r_1\left[ \frac{T_A}{1keV}\right]^{r_2},~~r_1\simeq 0.256~~,~~r_2\simeq 0.153 
\label{quench1}
\eeq 
Then the parameter $Qu(A)$ takes the values:
\beq
0.49,~0.38,~0.35,~0.31,~0.29\mbox{ for He, Ne, Ar, Kr and Xe}
\label{quench2}
\eeq
respectively. The effect of quenching is larger in the case of  heavy targets, since, for a given neutrino energy, the energy of
the recoiling nucleus is smaller. Thus the number of expected events for Xe assuming a threshold energy of
$100$ eV is reduced to about 560.\\
  The effect of quenching is exhibited in Fig \ref{fig:Kqu}  for the two interesting targets 
Ar and Xe.
 \\We should mention that it is of paramount importance to experimentally measure the quenching factor. The
 above estimates were based on the assumption of a pure gas. In our detection scheme the Xe gas carrier
 (A) is mixed with a small fraction of low ionization potential gas (B). Thus a part of the excitations
 produced on the Xe atoms could be transferred to ionization through the well known Penning effect
 as follows:
 \beq
 A^*+B\longrightarrow A+B^{*+}+e^{-}
 \label{penning}
 \eeq 
 Such an effect will lead to an increase  in the quenching factor and needs be measured.
\section{\label{sec:level6}Conclusions}
 In the present study it has been shown that it is quite simple to detect typical supernova
neutrinos in our galaxy, provided that such a supernova explosion takes place (one explosion every 30 years is estimated \cite{SOLBERG}). The idea is to employ a small size spherical TPC detector filled with a high
pressure noble gas. An enhancement of the neutral current component is achieved via the coherent
effect of all neutrons in the target. Thus employing, e.g., Xe at $10$ Atm, with a feasible threshold energy
of about $100$ eV in the detection the recoiling nuclei,
 one expects between $600$ and $1900$ events, depending on the quenching factor.
We believe that networks of such dedicated detectors, made out of simple, robust and cheap technology,
 can be simply managed by an international scientific consortium and operated by students. This network
 comprises a system, which can be maintained
for several decades (or even centuries). This is   is a key point towards being able to observe
 few galactic supernova explosions.
 
acknowledgments: This work was supported in part by the
European Union under the contracts
MRTN-CT-2004-503369 and the program PYTHAGORAS-1. The latter is part of the
Operational Program for Education and Initial Vocational Training of the
Hellenic Ministry of Education under the 3rd Community Support Framework
and the European Social Fund. One of the authors (JDV) is indebted for support and hospitality to the NANP05 organizing 
committee during the NANP05 conference and to Professor Hiroshi Toki of RCNP during the preparation of the manuscript.

\end{document}